\documentclass[aps,twocolumn,showpacs]{revtex4}
\usepackage{graphicx}

\begin{document}

\title{
High-Frequency  Properties of a  Graphene Nanoribbon
Field-Effect Transistor
}
\author{M.~Ryzhii, A.~Satou, and 
V.~Ryzhii 
}
\affiliation{
Computational Nanoelectronics Laboratory, University of Aizu, 
Aizu-Wakamatsu, 965-8580, Japan and\\
Japan Science and Technology Agency, CREST, Tokyo 107-0075, Japan
}
\author{T.~Otsuji}
\affiliation{Research Institute for Electrical Communication,
Tohoku University,  Sendai,  980-8577, Japan and\\
Japan Science and Technology Agency, CREST, Tokyo 107-0075, Japan
}
\date{\today}
\begin{abstract}
We propose an analytical device model for  a graphene nanoribbon
field-effect transistor (GNR-FET).
The GNR-FET under consideration is based on
 a heterostructure which consists of
an array of nanoribbons clad between the highly conducting
substrate (the back gate) and the top gate controlling
the dc and ac source-drain currents. 
Using the model developed,
we  derive  explicit analytical formulas
for the GNR-FET transconductance as a function of the signal frequency,
collision frequency of electrons, and the top gate length.
The transition from the ballistic and 
to strongly collisional electron transport
is considered.
\end{abstract}

\maketitle
\newpage

\section{Introduction}

The utilization of the patterned graphene which constitutes an
array of sufficiently narrow graphene nanoribbons as the basis
for field-effect transistors  (GNR-FETs)
opens up wide
opportunities  for the  energy band engineering promoting the achievement 
of  the optimal device parameters.
By properly choosing the width of the nanoribbons,
one can fabricate the graphene structures with relatively wide
band gap but rather high electron (hole) mobility~\cite{1,2}
(see also Refs.~\cite{3,4,5,6,7} and the references therein).

In this paper, we present an analytical device model for  a
graphene-nanoribbon FET (GNR-FET) and obtain the device characteristics. 
The GNR-FET under consideration
is based on a patterned graphene layer
which constitutes a dense  array of parallel
nanoribbons of width $d$ with the spacing
between the nanoribbons $d_s \ll d$. 
The nanoribbon edges are connected to the conducting pads
serving as the transistor source and drain.  
A highly conducting substrate plays the role of the back gate,
whereas 
the top gate serves to control the source-drain current.
The device structure is schematically 
 shown in Fig.~1.
Using the developed model, we calculate
the potential distributions
in 
the GNR-FET 
as a function of the back gate, top gate, and drain voltages,
$V_b$, $V_g$, and $V_d$ (reckoned from  the potential
of the source contact), respectively,
and the GNR-FET
dc  characteristics.
This corresponds to a GNR-FET in the common-source circuit.
The case of common drain will briefly be discussed as well.
For the sake of definiteness, the back gate voltage $V_b$ and
the top gate voltage $V_{g}$
are assumed to be, respectively,  positive and negative ($V_b > 0$
and $V_{g} < 0$) with respect to the potential
of the source contact, so we consider GNR-FETs with the channel of  n-type.

The paper is organized as follows. In Sec.~II,
the GNR-FET device model is considered.
Here the main equations governing the GNR-FET operation
are presented. 
Section~III deals with 
the derivation of the GNR-FET steady-state  characteristics.
The results of this section generalize those obtained previously~\cite{7}
for the case of ballistic electron transport
for  arbitrary values of the electron collision frequency.
In Sec.~IV, we derive a general formula for the ac source-drain current
as a function of the signal frequency and 
the device  material and structural
parameters.
Section~V is devoted to the analysis of the GNR-FET transconductance 
in a wide range of the signal frequencies and device parameters.
Section~VI deals with the interpretation of the obtained characteristics
and discussion of their limitations.
In Sec.~VII, we draw the main results.  
\begin{figure}[ht]
\begin{center}
\includegraphics[width=7.0cm]{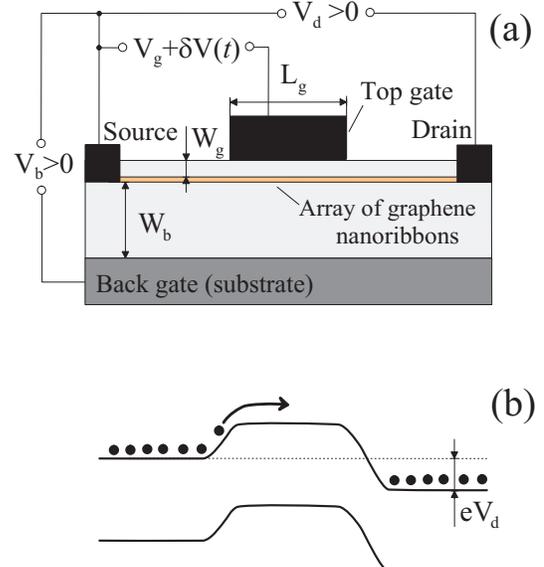}
\end{center}
\addvspace{-1 cm}\caption{ Schematic view  of (a)  a GNR-FET structure
and (b) its band diagram at the applied voltages.
}  
\end{figure}

\section{GNR-FET model}

We consider a GNR-FET with the structure shown in Fig.~1(a).
The back gate and drain voltages $V_b$ and $V_d$  (with $V_d < V_g$)
are  positive. As a result, the electron densities in the channel
sections adjacent  
to the sourse and drain (not covered by the top gate) are, respectively,
given by~\cite{7}
\begin{equation}\label{eq1}
\Sigma^s = \frac{\ae\,V_b}{4\pi\,eW_b}, 
\qquad \Sigma^d = \frac{\ae\,(V_b - V_d)}{4\pi\,eW_b},
\end{equation}
where $\ae$ is the dielectric constant, $W_b$ is the thickness
of the layer between the GNR array and the back gate.
The top gate voltage comprises the dc  and ac components, $V_g$ and $\delta\,V(t)$,
respectively ($V_g < 0, |\delta\,V(t)| \ll |V_g|$).
When the absolute value of the
dc component of the top gate voltage $V_g$ is sufficiently 
large ($|V_g| > V_bWg/W_b$),
the channel section beneath the top gate is essentially depleted
and potential barriers for the electrons incident from the source and drain sections
of the channel are formed (see Fig.~1(b)).
In the case of  a GNR-FET with sufficiently long top gate 
($L_g \gg \Lambda = \sqrt{W_bW_g/3}$), the height of the barriers
can be presented as~\cite{7}
\begin{equation}\label{eq2}
\Delta^{s,d}(t) = \Delta_0^{s,d} + e\delta\,V(t)\frac{W_b}{(W_b + W_g)},  
\end{equation} 
where $\Delta_0^s = - e( V_gW_b + V_bW_g)/(W_b + W_g)$
and   $\Delta_0^d = \Delta_0^s + eV_d$.
As seen from Eq.~(2), the variation of the top gate voltage leads to the variarion
of the potential barrier heights and, consequetly, to the variation
of the electron current above the barrier.

We assume that  the electron energy spectrum
of  the lowest  GNR subband can be presented as
\begin{equation}\label{eq3}
\varepsilon_{p} = 
\frac{\Delta}{2}\biggl[\sqrt{1 + \frac{2p^2}{\Delta\,m^*}} - 1\biggr].
\end{equation} 
Here the energy is reconed from the subband bottom,
$\Delta$ is the energy gap, $p$ is the momentum in 
along the nanoribbon, $m^* = \Delta/2v_W^2$
is the electron effective mass (near the subband bottom), and
 $v_W \simeq 10^8$~cm/s is the characteristic velocity of electrons
in graphene (and in GNRs). The energy gap $\Delta$ and the effective mass $m^*$
depend on the GNR widths $d$. In the simplest model, $\Delta \propto m^*
\propto d^{-1}$.
The electron transport in the section of the electron channel
beneath the top gate (the gated section), in which the electron density is small,
so that one can neglect the electron-electron scattering, can be described by
the following kinetic equation for the electron distribution function:

\begin{equation}\label{eq4}
\frac{\partial\, f(p,x,t)}{\partial\,t} + v_p\frac{\partial\,f(p,x,t) }{d\,x}
= -\nu_p[f(p,x,t) - f(- p,x,t)].
\end{equation}
Here 
\begin{equation}\label{eq5}
 v_p = \frac{d\,\varepsilon_p}{d\,p} = 
\frac{p}{m^*\displaystyle\sqrt{1 + \frac{2p^2}{\Delta\,m^*}}}.
\end{equation}
is the electron  velocity with the momentum $p$,
$\nu_p$ is the collision frequency of electrons associated with 
the disorder (including edge roughnesses) and acoustic phonons. We disregard the inelasticity of
the scattering on acoustic phonons, so that all the scattering
mechanisms under consideration result in the change 
of the electron momentum from
$p$ to $-p$. The probability of the electron elastic scattering is essentially
determined by the density of state. Taking into account collisional
broadening~\cite{8,9} (see also, Refs.~\cite{10,11}, we approximate the dependence in question as
$\nu_p = \nu\sqrt{\gamma^2 + 1}/\sqrt{\gamma^2 + (p/p_T)^2}$, 
where $\nu$ is the  value of the collisional frequency
of thermal electrons,
$\gamma$ chracterizes the collision broadening, 
and $p_T = \sqrt{2k_BTm^*}$.
If $|p|/p_T < \gamma$,  
$\nu_p \simeq \nu\sqrt{\gamma^2 + 1}/\gamma = const$, 
whereas at  $|p|/p_T \gg 1$ 
(but $|p| < p_{\Delta} = \sqrt{2\Delta\,m^*}$), one obtains
$\nu_p \simeq  \nu\sqrt{\gamma^2 + 1}(p_T/|p|)$.

The bondary conditions for $f(p,x,t)$ are set
at the points in the electron channel beneath the edges of the top gate, i.e.,
at $x = 0$ and $x = L_g$:
$$
f(p,x,t)|_{p \geq 0, x = 0} = f_M^s(p)
\exp\biggl[-\frac{\Delta^s(t)}{k_BT}\biggr],
$$
\begin{equation}\label{eq6}
f(p,x,t)|_{p \leq 0, x = L_g} = f_M^d(p)
\exp\biggl[- \frac{\Delta^d(t)}{k_BT}\biggr].
\end{equation}
Here $$
f_M^s(p) \propto \Sigma^s\exp\biggl(- \frac{\varepsilon_p}{k_BT}\biggr)
 \propto V_b\exp\biggl(- \frac{\varepsilon_p}{k_BT}\biggr),
$$
$$ 
f_M^d(p)\propto \Sigma^d\exp\biggl(- \frac{\varepsilon_p}{k_BT}\biggr)
\propto (V_b - V_d)\exp\biggl(- \frac{\varepsilon_p}{k_BT}\biggr).
$$
are the distribution functions of
the electrons incoming to the channel section in question from
the source and drain sides, respectively,
$k_B$ is the Boltzmann constant, and $T$ is the temperature,

\section{Steady-State Characteristics}

Solwing Eq.~(4) with boundary conditions~(6) at $\delta V_g (t) = 0$,
for the dc component of the electron distribution function
$f_0(p,x)$ we obtain the following relationship:
\begin{widetext}
$$
f_0(p,x) - f_0(-p,x) = \frac{1}{(1 + \beta_p)}
\biggl[f_M^s(p)
\exp\biggl(-\frac{\Delta_0^s}{k_BT}\biggr)
- f_M^d(p)
\exp\biggl(-\frac{\Delta_0^d}{k_BT}\biggr)\biggr]
$$
\begin{equation}\label{eq7}
= \frac{f^s_M(p)}{(1 + \beta_p)}
\exp\biggl[ \frac{e( V_gW_b + V_bW_g)}{(W_b + W_g)}\biggr]
\biggl[1 - \frac{(V_b - V_d)}{V_b}
\exp\biggl(- \frac{ eV_d}{k_BT}\biggr)\biggr].
\end{equation}
\end{widetext}
Here $\beta_p = \nu_pL_g/v_p$. For definiteness, here and in the following we assume
that the modulus of the top gate voltage $|V_g|$ is sufficiently large,
so that the channel beneath this gate is essentially depleted~\cite{7}.
At such top gate voltages, the GNR-FET dc transconductance is much larger that
when  $|V_g|$ is small.

The source-drain dc current can be calculated using
the following formula:
\begin{equation}\label{eq8}
J_0(x) = 
e\int_0^{\infty}dp\,v_p\, [f_0(p,x) - f_0(-p,x)] = J_0 = const.
\end{equation}
Considering Eqs.~(7) and (8), we obtain

$$
J_0 = e\biggl[\int_0^{\infty}dp\,v_p\,\frac{f^s_M(p)}{(1 + \beta_p)}\biggr]
\exp\biggl[\frac{ e( V_bW_g + V_gW_b)}{(W_b + W_g)k_BT}\biggr]
$$
\begin{equation}\label{eq9}
\times\biggl[1 - \frac{(V_b - V_d)}{V_b}
\exp\biggl(- \frac{ eV_d}{k_BT}\biggr)\biggr]
\end{equation}

As follows from Eq.~(3),
\begin{equation}\label{eq10}
v(\varepsilon_p) = \sqrt{\frac{2\varepsilon_p(\varepsilon_p + \Delta)}
{m^*\Delta[1 + 4\varepsilon_p(\varepsilon_p + \Delta)/\Delta^2]}}. 
\end{equation}
Since the kinetic energy of electrons propagating in the gated section of the channel
of realistic GNR-FETs
can be assumed small in comparison with the energy gap ($\varepsilon_p \sim k_BT \ll \Delta$), from Eq.~(11) one obtains the following simple relation: 
$v(\varepsilon_p) \simeq \sqrt{2\varepsilon_p/m^*}$. 
Taking this into account, from Eq.~(9) we obtain
\begin{equation}\label{eq11}
J_0 = J_0^B\,G_0.
\end{equation}
Here
$$
J_0^B = 
\biggl(\frac{ v_W\ae\,V_b}{2\pi^{3/2}}\biggr)\sqrt{\frac{k_BT}{\Delta}}
\exp\biggl[\frac{ e( V_bW_g + V_gW_b)}{(W_b + W_g)k_BT}\biggr]
$$
\begin{equation}\label{eq12}
= \biggl(\frac{\ae\,V_b}{2^{3/2}\pi^{3/2}}\biggr)\sqrt{\frac{k_BT}{m^*}}
\exp\biggl[\frac{ e( V_bW_g + V_gW_b)}{(W_b + W_g)k_BT}\biggr],
\end{equation}
is the dc source-drain current in the ballistic regime~\cite{7},
and the factor
$$
G_0 = 
\int_0^{\infty}\frac{d\xi\,\exp(- \xi)\sqrt{\xi}\sqrt{\gamma^2 + \xi}}
{(\sqrt{\xi}\sqrt{\gamma^2 + \xi} + \beta\sqrt{\gamma^2 + 1})}
$$
\begin{equation}\label{eq13}
= 1 -
\beta\sqrt{\gamma^2 + 1}\int_0^{\infty}\frac{d\xi\,\exp(- \xi)}
{(\sqrt{\xi}\sqrt{\gamma^2 + \xi} + \beta\sqrt{\gamma^2 + 1})}
\end{equation}
describes the effect of electron collisions.
Here $\beta = \nu\,L_g\sqrt{m^*/2k_BT}$ 
can be called the ballistic 
parameter
This parameter can also be presented as $\beta = \nu\tau$,  
where $\tau = L_g\sqrt{m^*/2k_BT}$ is the characteristic electron
transit time beneath the top gate and 
$v_T\sqrt{m^*/2k_BT}$ is the thermal electron velocity. 
The case $\beta \ll 1$ corresponds to near ballistic transport
of the majority of electrons (except, possibly, a fraction 
of low energy electrons). In the opposite case $(\beta \gg 1)$,
the  collisions substantially affect the electron transport.
When $\beta$ tends to zero,
$G_0$ turns to unity. 
Hence, in the ballistic regime of the electron transport through 
the section of the channel covered by the top gate, 
 Eq.~(11) yields $J_0 =J_0^B$, where the value $J_0^B$ as a function
of the gate and drain voltages
coincides 
with that obtained previously~\cite{7}.
In the case of collision-dominated electron transport $(\beta \gg 1)$,
from Eq.~(13)  we obtain
\begin{equation}\label{eq14}
G_0 \simeq \frac{a}{\beta},
\end{equation}
where $a$ is a coefficient (weakly dependent on $\gamma$).

Equations~(11) - (14) 
desribe the GNR-FET dc current-voltage characteristics,
i.e., the dependences of the dc source-drain current on the gate voltages
as well as the drain voltage, which generalize those derived 
previously~\cite{7}
for the purely ballistic transport.

\begin{figure}[ht]
\begin{center}
\includegraphics[width=7.0cm]{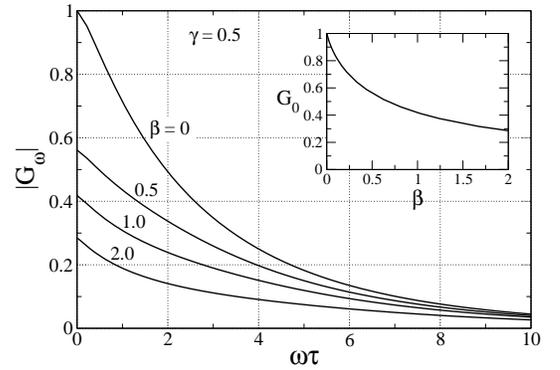}
\end{center}
\addvspace{0 cm}\caption{ Frequency dependences of the GNR-FET transconductance
modulus
for fifferent values of  ballistic parameter $\beta$.
Inset shows $G_0$ as a funcion of this parameter.
}  
\end{figure}

\begin{figure}[ht]
\begin{center}
\includegraphics[width=7.0cm]{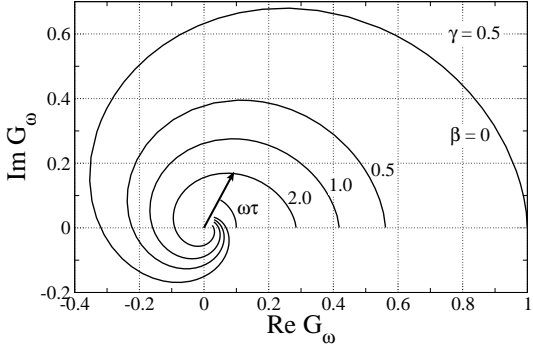}
\end{center}
\addvspace{0 cm}\caption{The GNR-FET amplitude-phase diagram for the same parameters as in Fig.~2. 
}  
\end{figure}

\section{Small-signal analisis}

Consider now the case when the top gate voltage comprises a small-signal 
comonent $\delta\,V(t) = \delta\,V_{\omega}\,\exp(-i\omega\,t)$, where 
$\delta\,V_{\omega}$
and $\omega$
are the signal amplitude and frequency.
In such a case, Eq.~(4) for the ac component of the distribution function
$\delta\,f_{\omega}(p,x)$ in the channel section beneath
the top gate reduces to the following:
\begin{equation}\label{eq15}
-i \omega \delta f_{\omega}(p,x) + v_p\frac{d\,\delta f_{\omega}(p,x) }{d\,x}
= -\nu_p[\delta f_{\omega}(p,x) - \delta f_{\omega}(- p,x)].
\end{equation}

As follows from Eq.~(6), the bondary conditions for $\delta\,f_{\omega}(p,x)$ 
in the most practical case  $eV_d \gg k_BT$ can be presented in the following form:
$$
\delta f_{\omega}(p,x)|_{p \geq 0, x = 0} = f_M^s(p)
\exp\biggl[\frac{ e( V_dW_g + V_gW_b)}{(W_b + W_g)k_BT}\biggr]\cdot\,
\frac{e\delta V_{\omega}}{k_BT},
$$
\begin{equation}\label{eq16}
\delta\,f_{\omega}(p,x)|_{p \leq 0, x = L_g} = 0.
\end{equation}

Solwing Eq.~(15) with boundary conditions (16) and using Eq.~(2), we arrive at the following
formula:
\begin{widetext}
$$
\delta f_{\omega}(p,x) - \delta f_{\omega}(- p,x) = 
2\frac{\omega\{(\tilde{\omega} + \omega)\exp[i\tilde{\omega}(x - L_g)/v_p]
+ (\tilde{\omega} - \omega)\exp[-i\tilde{\omega}(x - L_g)/v_p]\}}
{(\tilde{\omega} + \omega)^2\exp(-i\tilde{\omega}L_g/v_p] - 
(\tilde{\omega} - \omega)^2
\exp(i\tilde{\omega}L_g/v_p)}
$$
\begin{equation}\label{eq17}
\times
 f_M^s(p)
\exp\biggl[\frac{ e( V_bW_g + V_gW_b)}{(W_b + W_g)k_BT}\biggr]
\frac{W_b}{(W_b + W_g)}\frac{e\delta V_{\omega}}{k_BT},
\end{equation} 
where $\tilde{\omega} = \sqrt{\omega(\omega + i2\nu_p)}$.
Taking into account that the ac component of the electron current 
$\delta\,J_{\omega}$ in the channel beneath the top gate
is given by
\begin{equation}\label{eq18}
\delta\,J_{\omega}(x) = e\int_0^{\infty}dp\,v_p
[\delta\,f_{\omega}(p,x) - \delta\,f_{\omega}(- p,x)]
\end{equation} 
and using Eq.~(20), we obtain

$$
\delta\,J_{\omega}(x) = 
\biggl\{\int_0^{\infty}dp\,v_p\, f_M^s(p)
\frac{\omega\{(\tilde{\omega} + \omega)\exp[i\tilde{\omega}(x - L_g)/v_p]
+ (\tilde{\omega} - \omega)\exp[-i\tilde{\omega}(x - L_g)/v_p]\}}
{(\tilde{\omega} + \omega)^2\exp(-i\tilde{\omega}L_g/v_p] - 
(\tilde{\omega} - \omega)^2
\exp(i\tilde{\omega}L_g/v_p)}\biggr\}
$$
\begin{equation}\label{eq19}
\times
\exp\biggl[\frac{ e( V_bW_g + V_gW_b)}{(W_b + W_g)k_BT}\biggr]
\frac{W_b}{(W_b + W_g)}\frac{2e^2\delta V_{\omega}}{k_BT}.
\end{equation} 

In particular, near the drain-side edge of the top gate ($x = L_g$),
Eq.~(19) yields
$$
\delta\,J_{\omega}(L_g) = 
\biggl\{\int_0^{\infty}dp\,v_p\, f_M^s(p)
\frac{\omega\tilde{\omega}}
{(\tilde{\omega} + \omega)^2\exp(-i\tilde{\omega}L_g/v_p] - 
(\tilde{\omega} - \omega)^2
\exp(i\tilde{\omega}L_g/v_p)}\biggr\}
$$
\begin{equation}\label{eq20}
\times
\exp\biggl[\frac{ e( V_bW_g + V_gW_b)}{(W_b + W_g)k_BT}\biggr]
\frac{W_b}{(W_b + W_g)}\frac{4e^2\delta V_{\omega}}{k_BT}.
\end{equation} 
\end{widetext}
Equation~(20) can be presented as
\begin{equation}\label{eq21}
\delta\,J_{\omega}(L_g) = J_0^B\frac{W_b}{(W_b + W_g)}
\frac{e\delta V_{\omega}}{k_BT}G_{\omega},
\end{equation}
where

\begin{equation}\label{eq22}
G_{\omega} =
4\int_0^{\infty}
\frac{d\xi\,e^{- \xi}\,\omega\,\tilde{\omega}}
{(\tilde{\omega} + \omega)^2e^{-i\tilde{\omega}\tau/\sqrt{\xi}} - 
(\tilde{\omega} - \omega)^2
e^{i\tilde{\omega}\tau/\sqrt{\xi}}},
\end{equation}
where $J_0^B$ is given by  Eq.~(12) and
 $\tilde{\omega} = \sqrt{\omega(\omega + i2\nu\sqrt{\gamma^2 + 1}/
\sqrt{\gamma^2 + \xi})}$.

In this case, setting $\nu = 0$ in Eq.~(22), we obtain
\begin{equation}\label{eq23}
G_{\omega} =
\int_0^{\infty}
d\xi\,\exp\biggl(i\frac{\omega\tau}{\sqrt{\xi}} - \xi\biggr).
\end{equation}
Naturally, at $\omega = 0$, Eq.~(23) yields $G_{\omega} = G_0 = 1$.

\section{Transconductance}

\begin{figure}[t]
\begin{center}
\includegraphics[width=7.0cm]{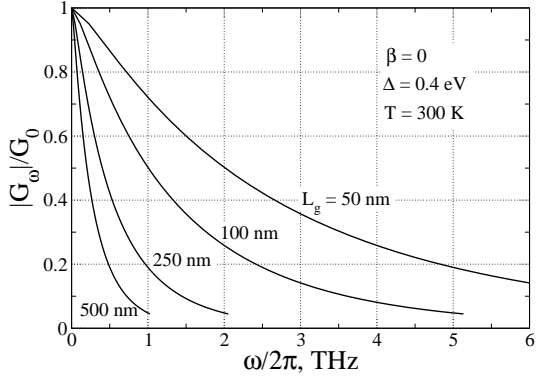}
\end{center}
\addvspace{0 cm}\caption{ Normalized transconductance vs signal frequency $f = \omega/2\pi$
calculated for ballistic ($\beta = 0$) GNR-FETs 
with different top gate lengths.
}  
\end{figure}

\begin{figure}[t]
\begin{center}
\includegraphics[width=7.0cm]{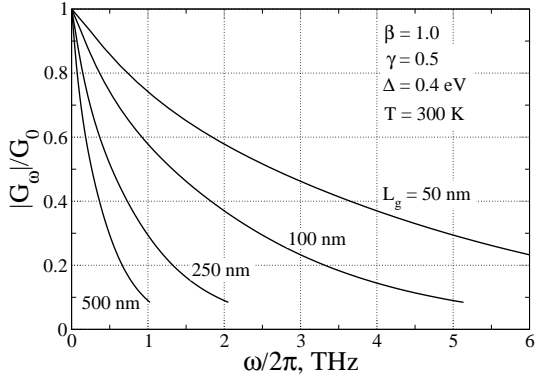}
\end{center}
\addvspace{0 cm}\caption{The same as in Fig.~4 but for GNR-FETs 
with essential electron scattering
($\beta = 1$).  
}  
\end{figure}
\begin{figure}[ht]
\begin{center}
\includegraphics[width=7.0cm]{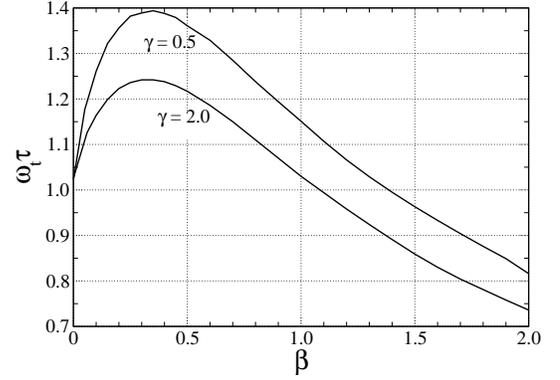}
\end{center}
\addvspace{0 cm}\caption{Normalized threshold frequency $\omega_t\tau$ 
vs ballistic parameter
$\beta$.  
}  
\end{figure}
The electrons passing from the the channel section beneath the top gate into
the drain section of the channel  between the edge of the top gate
($x = L_g$)
and the edge of the highly conducting
regions adjucent to the drain contact ($x = L_g + L_d$),  
induce the current through  the latter.
This induced source-drain terminal ac current can be calculated using 
the Shockley-Ramo theorem~\cite{12,13}
accounting for the features of the geometry of all highly conducting 
contacts~\cite{14,15}.  
The 
terminal ac current $\delta\,J_{\omega}$ is given by
\begin{equation}\label{eq24}
\delta\,J_{\omega} = \delta\,J_{\omega}(L_g)D_{\omega} = 
J_0^B\frac{W_b}{(W_b + W_g)}
\frac{e\delta V_{\omega}}{k_BT}G_{\omega}D_{\omega}.
\end{equation} 
Here $L_d$ is the length of the depleted region between the top gate
and the drain (see Fig.~1) determined by the applied dc voltages. 
In particular, if $V_d > V_b$, this depleted region extends to the 
drain contact. The electron drain transit-time  factor $D_{\omega}$
can be calculated using the following formula: 
\begin{equation}\label{eq25}
D_{\omega }
= 
\frac{1}{L_{d}}\int_{L_g}^{L_g + L_{d}} 
dx\,g(x)\exp\biggl[\frac{i\,\omega\tau_d\,(x - L_g)}{L_{d}}\biggr],
\end{equation}
where $\tau_d$ is the electron transit time across the gate-drain region,
 and $g(x)$
is the form-factor accounting for the shape of the highly conducting
regions (top gate and drain contact as well as the higly conducting
portion of the channel near the latter contact).
The electric field in the 
gate-drain region, so that their average velocity can substantially
exceed the electron thermal velocity. Hence, one might expect
that the electron transit time across the gate-drain region is
sufficiently short $\tau_d \ll \tau$. As a result, in the most interesting 
frequency range $\omega\tau \lesssim 1$, one obtains $\omega\tau_d \ll 1$.
In this case, one can put
$D_{\omega} = 1$. The transit-time effects which lead to
a significant frequency dependence of $D_{\omega}$
and, in particular, to a marked decrease in $|D_{\omega}|$
at elevated frequencies are discussed
in the following.

Figure~2 shows the frequency dependences of the GNR-FET transconductance
modulus
$|{\cal G}_{\omega}| = |G_{\omega}D_{\omega}| \simeq |G_{\omega}|$
calculated for different values of the ballistic parameter $\beta$.
It was assumed that $\gamma = 0.5$. The inset shows $G_0$
as a function of $\beta$. As it was demonstrated the $G_0$ vs $\beta$
dependence is almost insensitive to variations of parameter $\gamma$;
the maximum deviation from the dependence shown in the inset in Fig.~2
was less than $3.5\%$ when $\gamma$ varied from $\gamma = 0$ to 2.
In this and the following figures, we set $\Delta = 0.4$~eV 
and $T = 300$~K. 
Figure~3 demonstrates the amplitude-phase diagram calculated 
for the same parameters
as in Fig.~2.
The normalized transconductance 
$|G_{\omega}|/G_0$ as a function of the signal
frequency $f = \omega/2\pi$ calculated for GNR-FETs with
different top gate lengths is show in Figs.~4 and 5.

Figure~6 shows the dependence of the normalized theshold frequency
$\omega_t\tau$ at which $|G_{\omega}|/G_0 = 1/\sqrt{2}$.
It is instructive that the obtained dependences exhibit maxima
at certain values of the ballistic parameter $\beta$ (not at $\beta = 0$
as one might expect).
Moreover, the height of these maxima
decreases with increasing parameter $\gamma$, which characterizes
the collisional broadening (see Sec.~II).

\begin{figure}[t]
\begin{center}
\includegraphics[width=7.0cm]{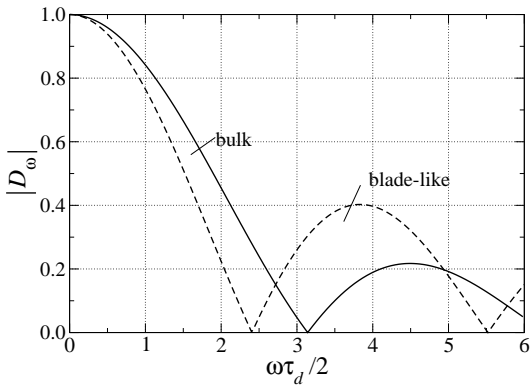}
\end{center}
\addvspace{0 cm}\caption{ Drain transit-time factor vs frequency: (a) for bulk drain contact
and (b) for blade-like drain contact. 
}  
\end{figure}

\section{Discussion}

The effect of the dependence
of the threshold frequency $\omega_t$
on parameters $\beta$ and $\gamma$
can be attributed to the following.
First, at the ballistic electron transport, the frequency dependence
of $G_{\omega}$ is actually determined not only by the characteristic
transit time $tau$ but also by the thermal spread in 
the electron velocities~\cite{16,17}. 
The latter is about the thermal electron
velocity $v_T $. The scattering of electrons, particularly strongly
dependent on the electron momentum, can influence the effective
 spead in velocities.
Indeed, when $\gamma \ll 1$, i.e., when the low-energy electrons
(with small $|p|$) exhibit rather strong scattering, their contribution
to the current decreases with simultanious decrease in the effective
 spead in velocities. 
 This can be seen, in particular, from Eq.~(7) in which
the asymmetric part of the distribution function $f_0(p,x)$
is 
$f_0(p,x) - f_0(-p,x) \propto f_M^s(p)/(1 + \beta_p)$. 
Since the term $\beta_p$ sharply increases when $p$ tends to zero,
this function can be fairly small in the range of small $|p|$ (especially
at small $\gamma$),
while at large $|p|$ it tends to the Maxwellian function.
This implies that, the momentun distribution can shrink and 
the velocity spread 
can  decrease due to specific feature of the electron scattering
in GNRs.
A significant increase in the threshold frequency can be achieved
if the directional electron velocity markedly exceeds
the thermal velocity. This can occur in GNR-FETs with the GNRs
with varied widths (increasing from the source-side edge of the top gate
to its drain-side) and the pertinent decrease in the energy gap.
In such hetero-GNRs, the electrons can accelerate propagating from
the source to drain (see, for instance, Ref.~\cite{18}).

As was mentioned above,  factor $D_{\omega}$ associated with the electron
transit-time effects in the region between the top gate and drain contact
depends on the transit angle $\omega\tau_d$ and the form-factor $g(x)$.
Figure~7 shows the frequency dependence of $|D_{\omega}|$ for the following
two cases: (a) $g(x) = 1$  and (b) $g(x) = 
L_d\sqrt{[x -(L_g -L_d/2)^2 - L_d^2/4]}$~\cite{14,15}.
The form-factor (a) can be attributed to the case when the drain contact
is rather bulk, so that the source-drain ac current is induced
primarily in this contact (not in the highly conducting portion of the 
channel adjacent to the drain contact). The form-factor (b) corresponds to
the case when the ac current is induced mostly in   the highly conducting
region or when the drain contact is a blade-like. 
Both such cases can take place depending on the spacing, $L_{gd}$, 
between the top
gate and the drain contact and the bias voltages $V_b$ and $V_d$. 
At $V_d > V_b$, the region of the channel between the top gate and
the drain contact is virtually depleted. 
Hence, if the drain contact is bulk,  $g(x) \simeq 1$ with $L_d = L_{sd}$.
The frequency dependences of $|D_{\omega}|$ calculated for
the cases under consideration  are shown in Fig.~7.
As seen from Fig.~7, $|D_{\omega}|$ markedly drops with increasing
transit angle $\omega\tau_d$ when the latter exceeds unity.
This effect is not important if the frequency-dependent
GNR-FET transconductance is mainly limited by the electron delay
under the top gate, i.e., if $\tau \gg \tau_d$.
However, at $\tau \lesssim \tau_d$, the net transconductance
is determined by the product of $|G_{\omega}|$ and $|D_{\omega}|$
corresponding to curves shown in Fig.~4 and Fig.~7, respectively.

\section{Conclusions}

(1) We proposed an analytical  device model for a  GNR-FET 
based on
 a heterostructure which consists of
an array of nanoribbons clad between the highly conducting
substrate (the back gate) and the top gate.

(2) Using this model, we derived  explicit analytical formulas
for the GNR-FET transconductance as a function of the signal frequency
in wide ranges of the 
collision frequency of electrons and the top gate length.

(3) The variation of the GNR-FET high-frequency
characteristics due to
the transition from the ballistic and to strongly collisional electron 
transport was traced. It was shown that the threshold frequency
can nonmonotonically depend on the ballsitic parameter.

\section{Acknowledgment}
The work was supported by CREST the Japan Science and 
Technology Agency, CREST,  Japan.

\newpage

\end{document}